\documentclass{appolb}
\usepackage{graphicx}

\begin{document}
\title{L\'evy-stable two-pion Bose-Einstein correlation functions measured with PHENIX in $\sqrt{s_{\textmd{NN}}} = 200$ GeV Au+Au collisions %
\thanks{XIIIth Workshop on Particle Correlations and Femtoscopy in Cracow, Poland}%
}
\author{S\'andor L\"ok\"os for the PHENIX Collaboration
\address{Eszterh\'azy University, M\'atrai \'ut 36, 3200 Gy{\"o}ngy{\"o}s, Hungary \\
E{\"o}tv{\"o}s University, P\'azm\'any P\'eter s\'et\'any 1/A, 1117 Budapest, Hungary}
}
\pagestyle{plain}
\maketitle
\begin{abstract}
Measurement of quantumstatistical correlation functions in high energy nuclear physics is an important tool to investigate the QCD phase diagram. It may be used to search for the critical point, and also to understand underlying processes such as in-medium mass modifications or partially coherent particle production. Furthermore, the measurements of the femtoscopic correlation functions shed light on the space-time structure of particle production. Consequently, the precise measurements and description of the correlation functions are essential. The shape of two-pion Bose-Einstein correlation functions were often assumed to be Gaussian, but the recent precision of the experiments reveals that the statistically correct assumption is the more general L\'evy-distribution. In this paper we present the recent results of the measurements of two-pion L\'evy-stable Bose-Einstein correlation functions in Au+Au collisions at PHENIX.
\end{abstract}
\PACS{25.75.-q, 25.75.Ag, 25.75.Dw}
  
\section{Introduction}

Bose-Einstein correlation measurements represent a broadly used technique in high energy nuclear physics. Intensity correlations were discovered in radioastronomy by R. Hanbury Brown and R. Q. Twiss, when they investigated the angular diameter of stars \cite{HBT_orig}. Independently, momentum correlations of identical pions were observed in proton-antiproton annihilation by Goldhaber and collaborators. This could be explained by the Bose-Einstein symmetrization of the pion wavefunction \cite{PhysRevLett.3.181}.

The Bose-Einstein correlations are related to the space-time distribution of the particle emitting source through a Fourier transform, hence the measured correlations have a clear connection to the size and the shape of the source. For the parameterization of this source usually a Gaussian profile was assumed, but in imaging measurements~\cite{Adler:2006as} a long tail was observed, motivating the use of the more general L\'evy distributions. In these proceedings paper we report about the two-pion Bose-Einstein correlations at $\sqrt{s_{\textmd{NN}}} = 200$ GeV Au+Au collisions in 0-30\% centrality \cite{Adare:2017vig}.

\section{The PHENIX experiment}

A detailed description of the PHENIX detector system can be found in Ref.\cite{Adcox:2003zm}. We reduce our discussion to the most important detectors used in this analysis. The Beam-Beam Counters (BBC) and Zero Degree Calorimeters (ZDC) were used to characterize the events. The Drift Chamber (DC) and the Pad Chambers (PC) were used for tracking. We used charged pions in the analysis which were identified with lead scintillators (PbSc) and high-resolution time-of-flight (ToF) detectors in both detector arms. We measured pions in the $0.2$ GeV/c $\leq p_t \leq 0.85$ GeV/c transverse momentum range.

\section{Two-particle correlations and the L\'evy distribution}

The two-particle correlation functions can be defined with the single-particle and pair momentum distributions as
\begin{equation}
C_2 = \frac{N_2(p_1,p_2)}{N_1(p_1)N_1(p_2)}.
\end{equation}
The momentum distribution can be expressed with the $S(x,p)$ source distribution function as~\cite{Yano:1978gk}
\begin{equation}
N_2(p_1,p_2)=\int dx_1^4 dx_2^4 S(x_1,p_1)S(x_2,p_2) \left| \Psi_{p_1,p_2}(x_1,x_2) \right|^2
\end{equation}
where $\Psi$ is the symmetrized pair wave-function. The one-particle momentum distribution provides a normalization
for this. The $C_2$ correlation function can then be written up with the source function as:
\begin{equation}
C_2(q,K)=1 + \left| \frac{\tilde{S}(q,K)}{\tilde{S}(0,K)} \right|^2
\end{equation}
where $q=p_1-p_2$ is the momentum difference and $K=0.5(p_1+p_2)$ is the average pair transverse momentum and $\tilde{\:}$ denotes the Fourier transform respect to the variable $x$. This above expression for $C_2$ takes the value of 2 at $q=0$ relative momentum, per definition. However, our (and in general, most high energy nuclear physics) measurements cannot resolve momentum differences below a few MeV/c, therefore we only can extrapolate the measured correlation functions to $q=0$. It turns out, that the extrapolated value of the $C_2$ does not reach the value 2, but $1+\lambda$. This observation is quantified with the intercept parameter as $0 < \lambda \leq 1$. The value of the intercept parameter $\lambda$ may be explained in term of the core-halo picture~\cite{Csorgo:1999sj}: it is related to the core fraction of the particle producing source as $\lambda = (N_{\rm core}/N_{\rm total})^2$ (where the core is surrounded by a halo of the decay products of long lived resonances), as detailed in Refs.~\cite{Csorgo:1999sj,Adare:2017vig}. With these, the correlation function can be given as
\begin{equation}
C_2(q,K) = 1+\lambda(K)\left| \frac{\tilde{S}(q,K)}{\tilde{S}(0,K)} \right|^2 \: \: \: \: \textmd{where} \: \: \: \: \lambda = \left( \frac{N_\textmd{core}}{N_\textmd{core}+N_\textmd{halo}} \right)^2.
\label{eq:C2}
\end{equation}

In heavy-ion experiments the shape of the above discussed correlation functions are usually assumed to be Gaussian. One may have to renounce this simple premise once the expansion of the source created in the collision is taken into account. In the expanding hadron gas the particles have an increasing mean-free path, which may lead to anomalous diffusion and the appearance of L\'evy distributions \cite{Csanad:2007fr}. The one-dimensional, symmetric L\'evy-distribution is defined by a Fourier transform as
\begin{equation}
\mathcal{L} = \frac{1}{(2\pi)^3}\int d^3\textbf{q} e^{i\textbf{q} \cdot \textbf{r}}e^{-\frac{1}{2}|\textbf{q}R|^\alpha}.
\label{eq:levy}
\end{equation}
This distribution has two parameters: the $\alpha$, so-called stability index, and the $R$ L\'evy scale or size parameter. If we assume that the source function has L\'evy-shape, with Eq. (\ref{eq:C2}) the following can be deduced:
\begin{equation}
C_2^{(0)}(q,K) = 1 + \lambda (K) e^{-(R(K)q)^{\alpha(K)}}.
\label{levycorr}
\end{equation}
The $^{(0)}$ index indicates that the none of the final state effects are taken into account. In our case the only important one is the Coulomb repulsion of the measured particles. We used the modified Sinyukov type of method as detailed in Ref.~\cite{Adare:2017vig,Sinyukov:1998fc}.

\section{Results}

We measured two-pion Bose-Einstein correlation functions and parametrized them with the above detailed L\'evy-type correlation function, as also discussed in Ref.~\cite{Adare:2017vig}. We determined the $m_T$ dependence of the parameters in 31 bins,
using a 0-30\% centrality selection in $\sqrt{s_{\textmd{NN}}} = 200$ GeV Au+Au collisions.

\begin{figure}[htb]
\centerline{
\includegraphics[width=0.49\textwidth]{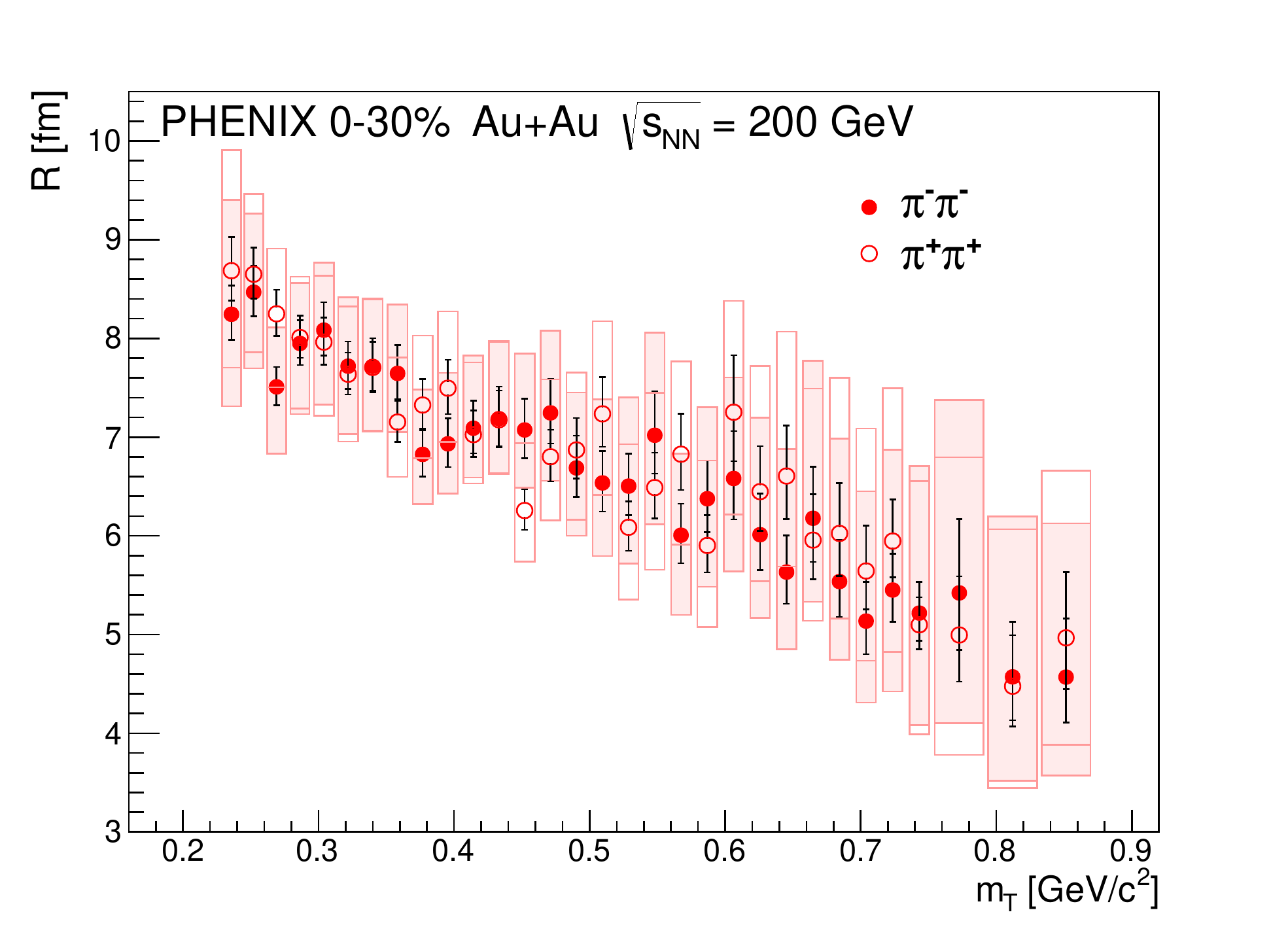}
\includegraphics[width=0.49\textwidth]{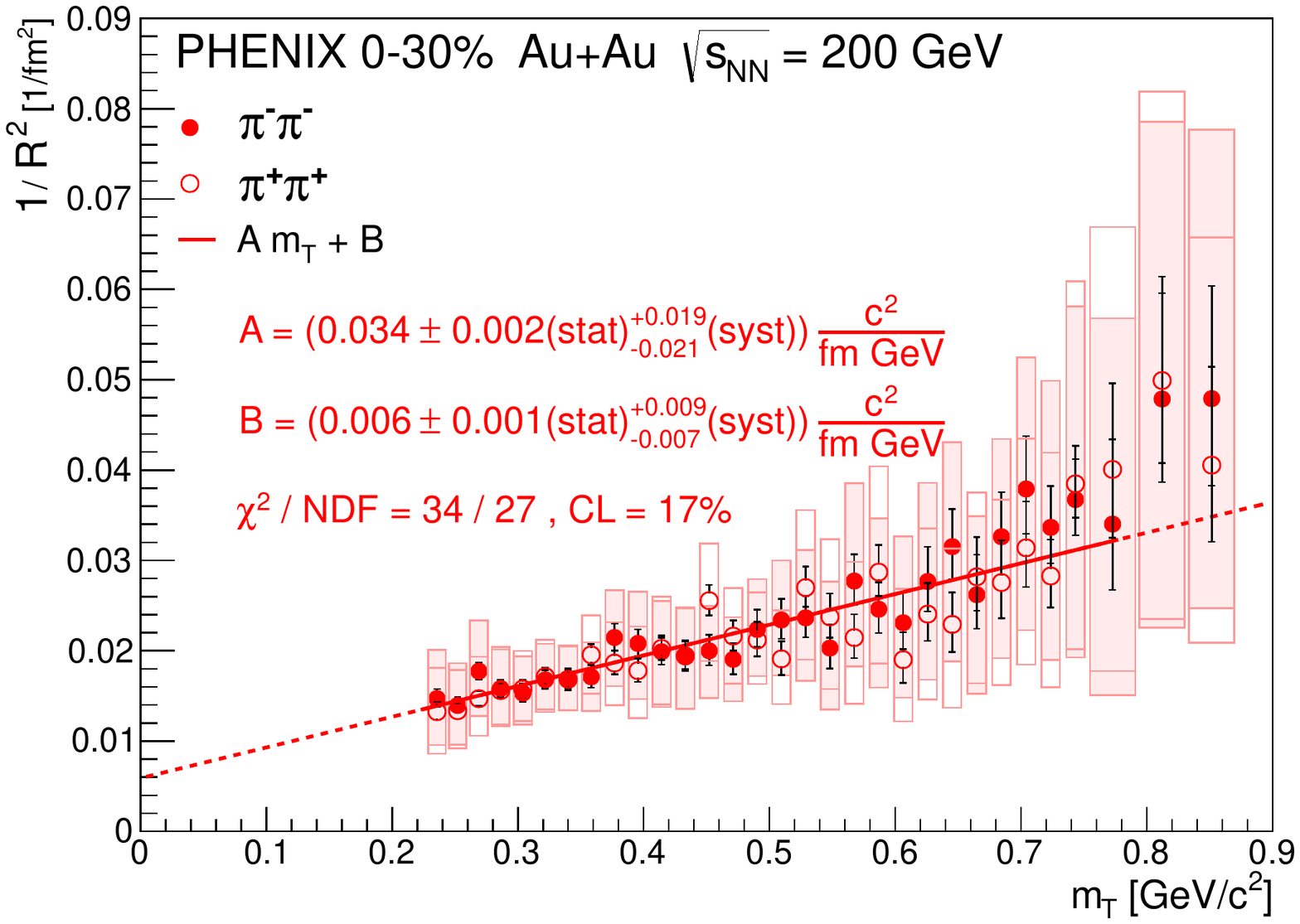}}
\caption{The L\'evy-scale (left hand side) shows similar trend as in the Gaussian case. The $1/R^2$ scaling behavior predicted from hydrodynamical model calculations e.g. in Ref.\cite{Makhlin1988,Csorgo:1995bi,Chapman:1994yv} also stay valid (right hand side).}
\label{fig:R}
\end{figure}

The obtained $R$ L\'evy scale parameter results are shown in Fig.~\ref{fig:R}. While these values may have to be interpreted differently from the usual Gaussian HBT radii, they shows similar, hydro inspired $R \propto 1/\sqrt{m_T}$, as expected from hydrodynamics. 

\begin{figure}[htb]
\centerline{
\includegraphics[width=0.49\textwidth]{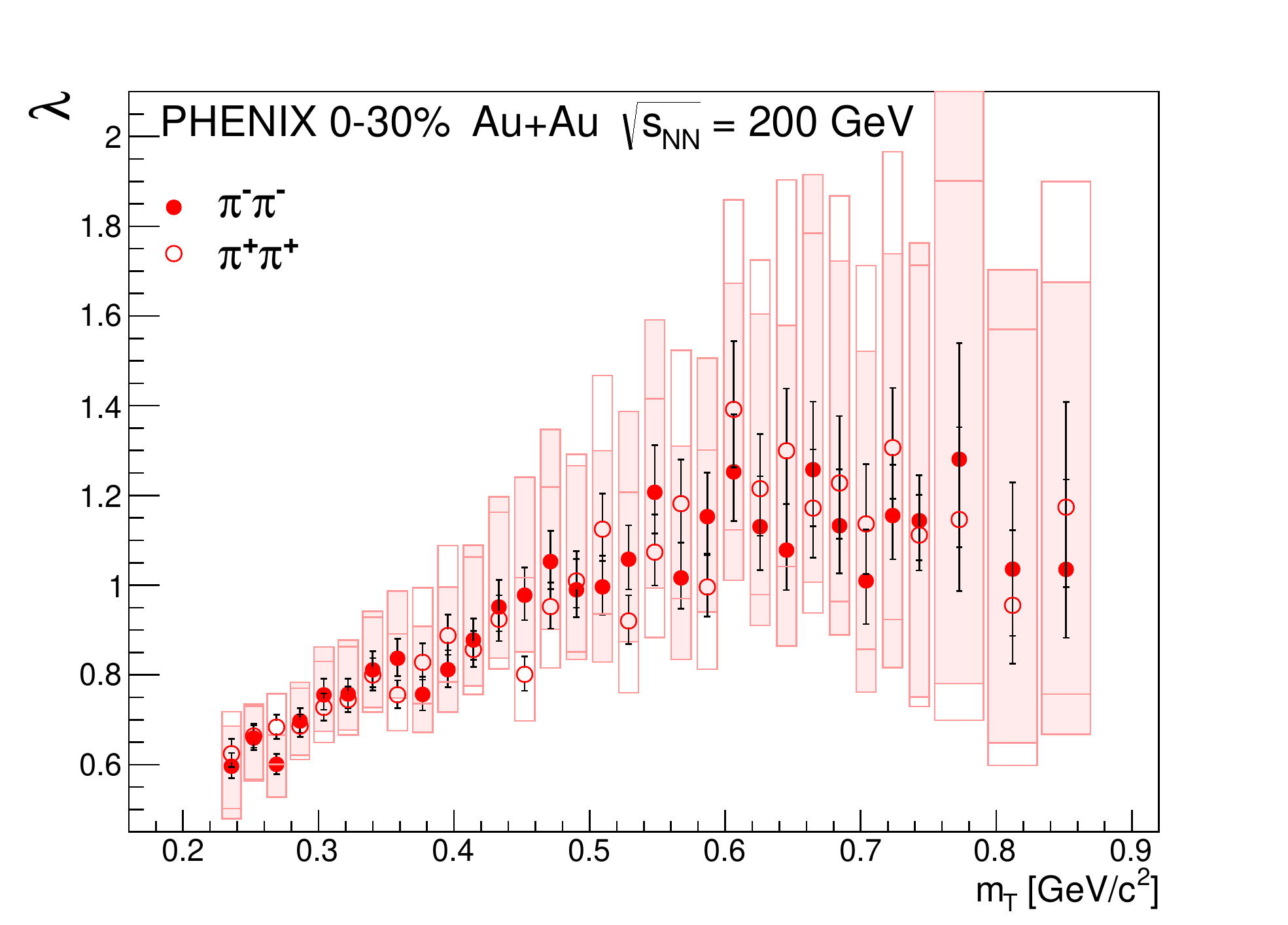}
\includegraphics[width=0.49\textwidth]{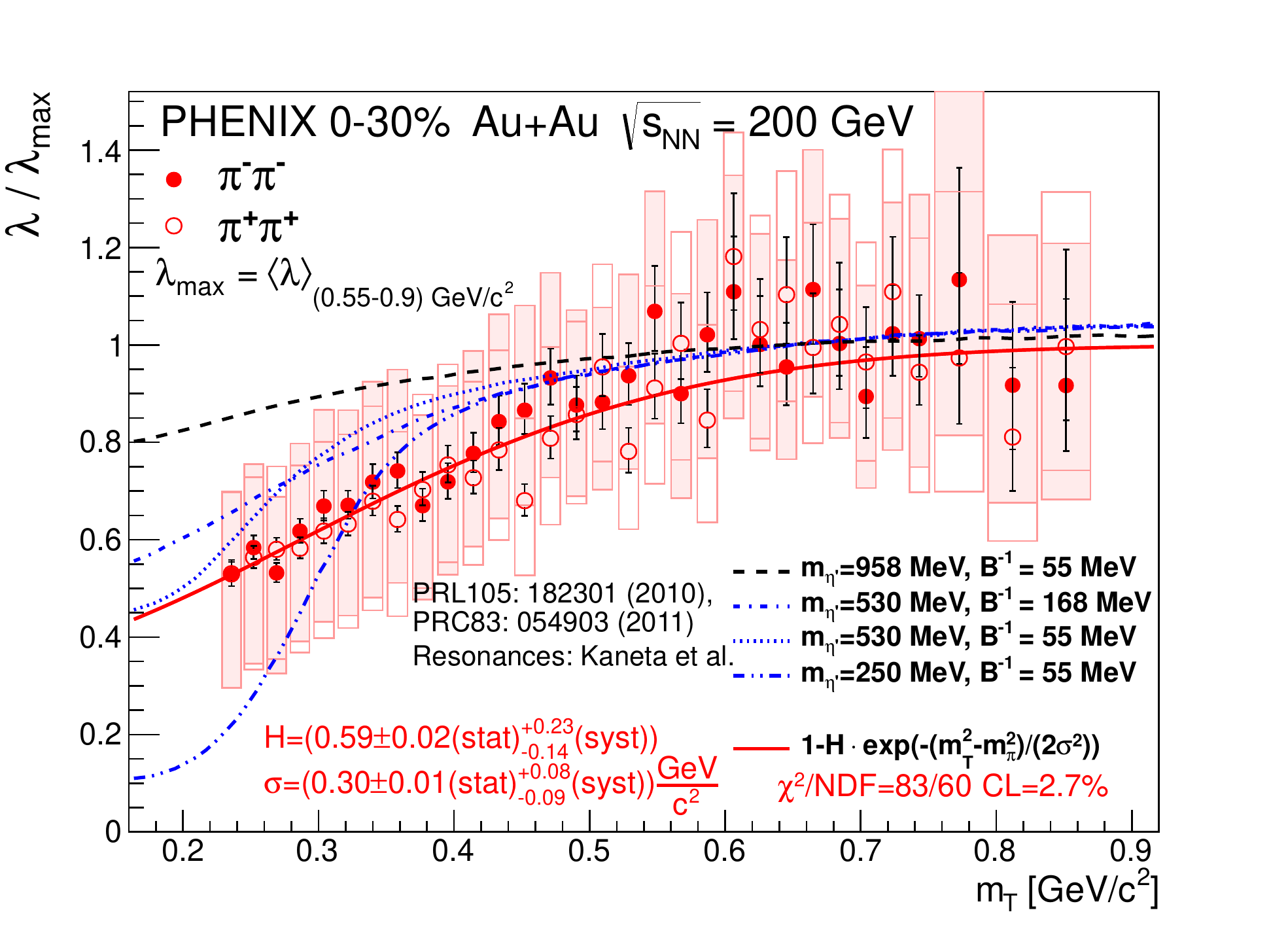}}
\caption{The measured intercept parameter (left hand side) and its normalized version (right hand side). In both plots the decreasing trend is clearly visible.}
\label{fig:lambda}
\end{figure}

In sufficiently hot and dense QCD matter, the anomalously broken $U_A(1)$ symmetry may be restored, in which case the $\eta'$ meson has reduced mass, hence more $\eta'$ mesons will be produced. The $\eta'$ decays also into pions, which contribute to the halo, hence decrease $\lambda$. Due to the specific kinematics, a low $m_T$ suppression was predicted, as detailed in Ref.~\cite{Vertesi:2009wf}.
The measured $\lambda(m_T)$ is not incompatible with this prediction, as indicated in the left plot of Fig.~\ref{fig:lambda}, where the ``hole'' or decreasing trend of the intercept parameter is clearly visible. On the right hand side the $\lambda/\lambda_{\textmd{max}}$ is presented along with a unity minus Gaussian fit and resonance model predictions.

\begin{figure}[htb]
\centerline{
\includegraphics[width=0.49\textwidth]{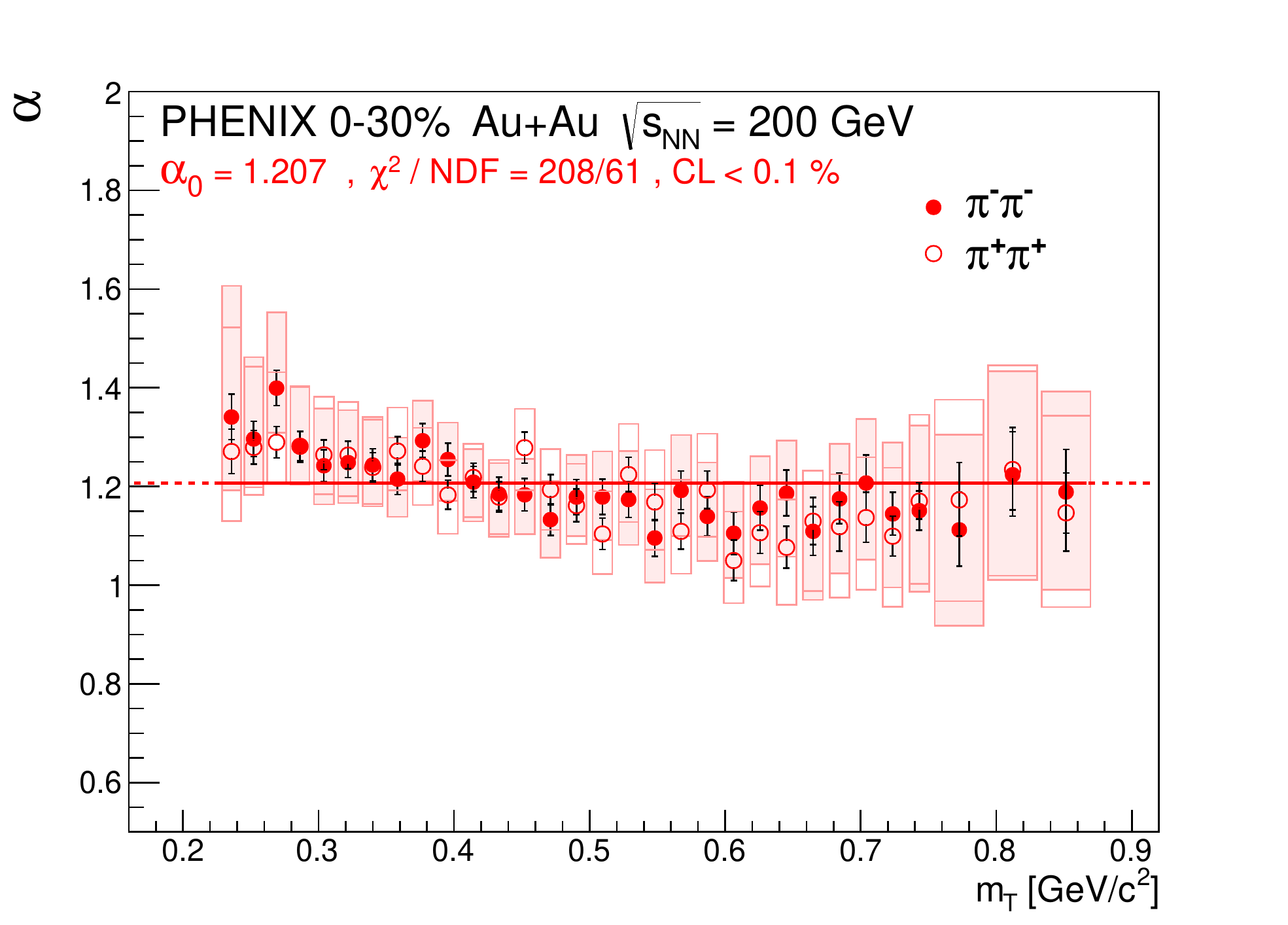}
\includegraphics[width=0.49\textwidth]{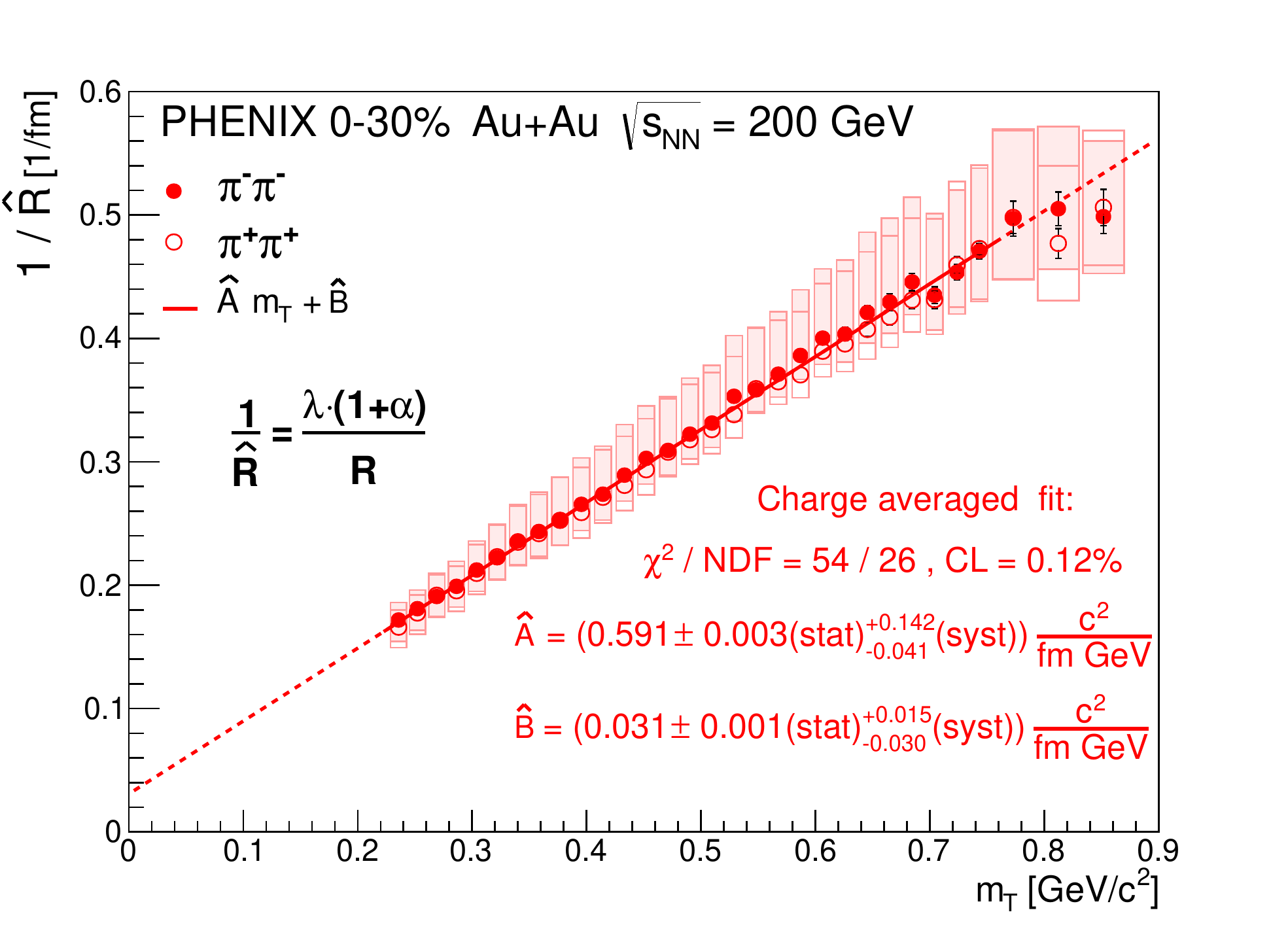}}
\caption{The L\'evy shape parameter $\alpha$ (left hand side) exhibits a slight non-monotonic behavior as a function of $m_T$ and has the average value differs from the Gaussian and the conjectured critical value. The new scaling parameter $\widehat{R}$ is remarkably linear as the function of $m_T$.}
\label{fig:alpha_rhat}
\end{figure}

The $\alpha$ shape parameter can characterize the deviation of the source $S(r)$ from the Gaussian or the Cauchy distributions. In case of $\alpha = 2$ the Gaussian case is restored, while $\alpha = 1$ corresponds to a Cauchy shaped source, and an exponential correlation function. This parameter is also sometimes associated to one of the critical exponents, namely to the critical exponent of the spatial correlations~\cite{Csanad:2007fr,Csorgo:2003uv}. Thus the precise measurements of this parameter in various systems could indicate the vicinity of the supposed critical point of the quark-hadron transition on the QCD phase diagram. Proceedings publications on the beam energy and centrality dependence of this parameter can be found in Refs.~\cite{Kincses:2017zlb,Lokos:2018dqq}. The left panel of fig.~\ref{fig:alpha_rhat} shows the measured $\alpha$ parameter as a function of average pair $m_T$. This figure shows that thee $\alpha$ parameter in 200 GeV Au+Au collisions is between the mentioned special cases (Gaussian and Cauchy) and has a slightly non-monotonicity as a function of $m_T$.

Finally, let us note that we found a new empirical scaling parameter which is composed from the three L\'evy parameters as
\begin{equation}
\frac{1}{\widehat{R}} = \frac{\lambda(1+\alpha)}{R}.
\end{equation}
Its value versus $m_T$ is shown in the right panel of Fig.~\ref{fig:alpha_rhat}. A very clear linear trend in $1/\widehat{R}$ versus $m_T$ can be observed, as well as a reduction of the statistical uncertainty. The latter can be explained by the correlation of the other fit parameters, and by $\widehat{R}$ being a ``strong mode'' of these L\'evy fits. However, the linear connection shown in Fig.~\ref{fig:alpha_rhat} is nor predicted neither explained by any of the known model as far as we know.

\section{Conclusions}

We measured two-pion L\'evy-stable Bose-Einstein correlation functions at $\sqrt{s_{\textmd{NN}}} = 200$ GeV in Au+Au collisions at PHENIX. We parametrized these with correlation functions calculated from a theoretically motivated generalization of the Gaussian distribution: the L\'evy-distribution. This yields a statistically acceptable description of the measured data. We determined the $m_T$ dependence of the L\'evy fit parameters. The L\'evy stability parameter is measured to be different from any previously assumed special distribution and has a weak $m_T$ dependence. We furthermore concluded that the measured values and trends of the parameters do not contradict the partial restoration of the $U_A(1)$ symmetry. We also observed a hydro-predicted scaling of the L\'evy scale versus $m_T$, as well as a new empirical scaling parameter $\widehat{R}$.

\section*{Acknowledgments}
The author was supported by the NKFIH grant FK 123842.

\end{document}